\documentclass{emulateapj}

\def\tobs{t_{\rm obs}}

\def\nuobs{\nu_{\rm obs}}

\def\be{\begin{equation}}
\def\ee{\end{equation}}
\def\beq{\begin{eqnarray}}
\def\eeq{\end{eqnarray}}

\begin{document}

\title{Towards an understanding of GRB prompt emission mechanism: \\
I. The origin of spectral lags}

\author{Z. Lucas Uhm, Bing Zhang}
\affil{Department of Physics and Astronomy, University of Nevada, Las Vegas, NV 89154, USA}

\begin{abstract}
Despite decades of investigations, the physical mechanism that powers the bright prompt $\gamma$-ray emission from gamma-ray bursts (GRBs) is still not identified. One important observational clue that remains not properly interpreted so far is the existence of time lags of broad light curve pulses in different energy bands, named ``spectral lags''. Here we show that the traditional view invoking the high-latitude emission ``curvature effect'' of a relativistic jet cannot account for spectral lags. Rather, the observed spectral lags demand the sweep of a spectral peak across the observing energy band in a specific manner. The duration of the broad pulses and inferred typical Lorentz factor of GRBs require that the emission region is in an optically thin emission region far from the GRB central engine. We construct a simple physical model invoking synchrotron radiation from a rapidly expanding outflow. We show that the observed spectral lags appear naturally in our model light-curves given that (1) the gamma-ray photon spectrum is curved (as observed), (2) the magnetic field strength in the emitting region decreases with radius as the region expands in space, and (3) the emission region itself undergoes rapid bulk acceleration as the prompt $\gamma$-rays are produced. These requirements are consistent with a Poynting-flux-dominated jet abruptly dissipating magnetic energy at a large distance from the engine. 
\end{abstract}


%
%

\section{Introduction}

Gamma-ray bursts (GRBs) are the most luminous explosions in the universe. They are produced by relativistic jets with the Lorentz factors of a few hundreds beaming toward Earth, making them the fastest moving astrophysical objects in the universe.  
Despite the extensive studies for many decades, however, the exact composition of the GRB jets, their energy dissipation mechanism, and the radiation processes involved, still remain unclear \citep[e.g.][for recent reviews]{kumarzhang15,peer15}. The traditional model suggests that the GRB jets are matter-dominated fireballs \citep{paczynski86,goodman86,shemi90} with $\gamma$-ray emission either originating from the internal shocks \citep{rees94} or the fireball photosphere \citep{meszarosrees00,rees05}. An alternative model invokes a Poynting flux dominated outflow from the central engine, with the emission region either matter dominated \citep[e.g.][]{drenkhahn02,thompson06,giannios08} or still moderately Poynting-flux-dominated \citep[e.g.][]{zhangyan11}.

Abundant prompt emission data have been accumulated over the years \citep[e.g.][for reviews]{fishman95,kumarzhang15}. 
Recently, extensive modeling of the observed spectra \citep[e.g.][]{meszarosrees00,peer06,beloborodov10,lazzati10,vurm11,daigne11,lundman13,uhm14} and light curves \citep[e.g.][]{hascoet12,zhangzhang14} has been carried out within the framework of different models, and detailed fitting to the data has been carried out for some of these models \citep[e.g.][]{burgess14,zhangbb15,ahlgren15}. However, due to the lack of smoking gun signatures, the case is so far inconclusive.

On the other hand, prompt emission observations have provided rich features that have not been modeled properly. These include, for example, the systematic time lags between light curves at different energy bands (the so-called ``spectral lags'', e.g. \citealt{norris96,norris00,liang06,lu15}), and the peak energy ($E_p$) evolution patterns across the (energy-dependent) pulse profiles \citep[e.g.][]{hakkila11,lu12}. In a series of papers, we plan to systematically model these observational features, as an effort towards an understanding of the GRB prompt emission mechanism.

This first paper aims at studying the physical origin of spectral lags. The GRB light curves typically show diverse and complex behaviors, with possible superpositions \citep{gao12} of fast variabilities on slowly varying ``pulses" \citep{norris96,hakkila11}. Already in the BATSE era, some early investigations suggested that the separated slow component shows an interesting spectral-lag behavior. The light curves at higher energies (say, in the MeV regime), in general, peak earlier and have a narrower width than those at lower energies (say, 10s to 100s of keV). Even though the quantitive fractions differ in different analyses, the dominant fraction of GRB pulses show such {\em positive} lags. A small fraction of pulses show {\em negative} lags (with the higher-energy emission slightly lagging behind the lower-energy emission) or even no spectral lags \citep[e.g.][]{norris96,norris00,liang06,ukwatta12,lu15}. Some attempts have been carried out to model these lags \citep[e.g.][]{ioka01,norris02,dermer04,shen05}. However, one could not simultaneously reproduce both the spectral lags and the peak-time fluxes of pulses in different energy bands. In this paper, we present a thorough investigation on the origin of spectral lags in GRBs, and derive the necessary physical conditions to interpret the observations.

%
%

\section{Curvature effect and the spectral lags}

Photons emitted from a relativistic spherical shell into the direction of a distant observer spread over the observer's time axis when received by the detector on Earth, even though they are emitted simultaneously in the co-moving frame of the shell. Due to the curvature of the spherical shell, emission from progressively higher latitudes with respect to the observer's line of sight is progressively delayed while being boosted with a smaller Dopper factor $D=[\Gamma (1-\beta \cos \theta)]^{-1}$. Here, $\Gamma$ denotes the Lorentz factor of the shell's bulk motion, $\beta$ is given by $\beta=(1-1/\Gamma^2)^{1/2}$, and the polar angle $\theta$ measures the latitude of the emission location with respect to the observer's line of sight. These aspects of the high-latitude emission are called the ``curvature effect.'' Based on such an effect, when the emission of the shell is turned off at a certain point, the light curves at lower energies (i.e., boosted with smaller Doppler factors) are expected to be delayed and exhibit a peak at a later time. For this reason, the curvature effect has long been suggested as an explanation for the spectral lags seen in the GRB prompt emission \citep[e.g.][]{ioka01,norris02,dermer04,shen05,shenoy13}. The main difficulty of this model, as appreciated by previous authors, is that even though the delay time scales may be reproduced, the flux levels from different energy bands are very difficult to reproduce. In particular, to interpret the observed time lags among different energy bands with the curvature effect model, the predicted fluxes at lower energy bands are significantly lower than what are observed (e.g. Section 4.6 of \citealt{zhang09} and references therein). In the following, we will show that this curvature effect model {\em cannot} give rise to spectral lags unless the spectrum is very narrow, and that even for the cases with narrow spectrum, the predicted light curves are very different from the observations.

\subsection{A simple physical model} \label{section:physical_model}

The physical picture adopted in our numerical model is simple and straightforward to understand: a single relativistic spherical shell expands radially with the bulk Lorentz factor $\Gamma$, continuously emitting photons from all locations in the shell, with an isotropic angular distribution of the emitted power in its co-moving fluid frame. The shape of the photon spectrum in the co-moving frame is delineated by a functional form \citep{uhm15a},
\be
\label{eq:Hx}
H(x)
\quad
\mbox{with}
\quad
x=\nu^{\prime}/\nu_{\rm ch}^{\prime}.
\ee
The $H(x)$ function is allowed to have an arbitrary shape as a function of the photon frequency $\nu^{\prime}$ (measured in the co-moving frame), and is located at a characteristic frequency $\nu_{\rm ch}^{\prime}$. The radiating electrons are distributed uniformly in the shell, and the number of the electrons $N$ is assumed to increase at an injection rate $R_{\rm inj} \equiv dN/dt^{\prime}$ from an initial value $N = 0$. Here, $t^{\prime}$ is the time, measured in the co-moving frame.

The luminosity distance to the GRB site is calculated adopting the standard flat $\Lambda$CDM universe with the parameters $H_0=71$ km $\mbox{s}^{-1}$ $\mbox{Mpc}^{-1}$, $\Omega_{\rm m}=0.27$, and $\Omega_{\Lambda}=0.73$. The redshift plays only a global role in shaping the observed spectral flux \citep{uhm15a}, and thus we assume a typical value $z=1$ in all numerical models presented in this paper.

The prompt gamma-rays of GRBs are not emitted from the explosion center, but rather at a certain radius far from the central engine. Hence, we consider that the emission of the spherical shell is turned on at radius $r_{\rm on}$ (and at the lab-frame time $t_{\rm on}$), such that the observer time $t_{\rm obs}$ is set equal to zero when the first photons from this radius $r_{\rm on}$ trigger the detector on Earth. Then, a photon emitted from the shell at a later lab time $t$ ($>t_{\rm on}$) when the shell has radius $r$ ($>r_{\rm on}$) will be received by the observer at
\beq
t_{\rm obs} &=&  \frac{1}{c} \left[r_{\rm on} + c(t-t_{\rm on}) - r \cos \theta \right] (1+z) \nonumber \\
&=& \left[ \left(t-\frac{r}{c} \cos \theta \right) - \left(t_{\rm on}-\frac{r_{\rm on}}{c}\right) \right] (1+z),
\eeq
where $c$ is the speed of light, and the angle $\theta$ measures the latitude of the photon's emission location with respect to the observer's line of sight. Notice that the observer time $\tobs$ is practically independent of $t_{\rm on}$ since we calculate the time $t$ as $t=t_{\rm on}+\int_{r_{\rm on}} dr/(c\beta)$.

We assign a spectral power $P_0^{\prime}$ to each electron in the shell and calculate the observed spectral flux, $F_{\nu_{\rm obs}}^{\,\rm obs}$, in terms of the observer time $t_{\rm obs}$ and the observed frequency $\nu_{\rm obs}$, while fully taking into account the high-latitude curvature effect of the spherical shell, which includes the Doppler boosting from the co-moving frame to the observer frame for each latitude and the integration over the equal-arrival time surface of each observer time \citep{uhm15a}.

We begin our calculations at radius $r_{\rm on}$ and turn off the emission of the shell at radius $r_{\rm off}$. As for the shape of the photon spectrum in the co-moving frame, we use a curved spectrum such as the Band function \citep{band93} with the parameters $\alpha_{\rm B}$ (the low-energy photon index) and $\beta_{\rm B}$ (the high-energy photon index). If the photon spectrum is not curved (say, a power-law shape), the light curves at different energies will always behave in the same way, rising and/or falling simultaneously, which then leaves no possibility to produce a spectral lag.

Our calculations are not attached to a particular theoretical model of prompt emission. In the following, we adopt a general approach with the goal to identify the physical conditions that are required by the observations.

\subsection{Curvature effect cannot interpret the observed spectral lags}\label{sec:curvature}

We first notice that GRB spectral lags sample the broad pulses in the light curves, which are the ``slow components'' in contrast to the fast-variability components that are superposed on the slow ones \citep{gao12}. The typical duration of the pulses is seconds. Given the typical Lorentz factor $\Gamma=100 \Gamma_2$ of GRBs, the typical emission radius is
\begin{equation}
 r \sim \Gamma^2 c\, t_{\rm pulse} \sim (3\times 10^{14} ~{\rm cm}) \Gamma_2^2 (t_{\rm pulse}/1~{\rm s}).
 \label{eq:r}
\end{equation}
Such a radius is much larger than the photosphere radius, suggesting that the emission occurs in an optically thin region above the photosphere. It is also consistent with the typical input parameters in our previous calculations to interpret the standard Band function spectra of GRBs within the fast-cooling synchrotron radiation model \citep{uhm14}.

\begin{figure}
\begin{center}
\includegraphics[width=18cm]{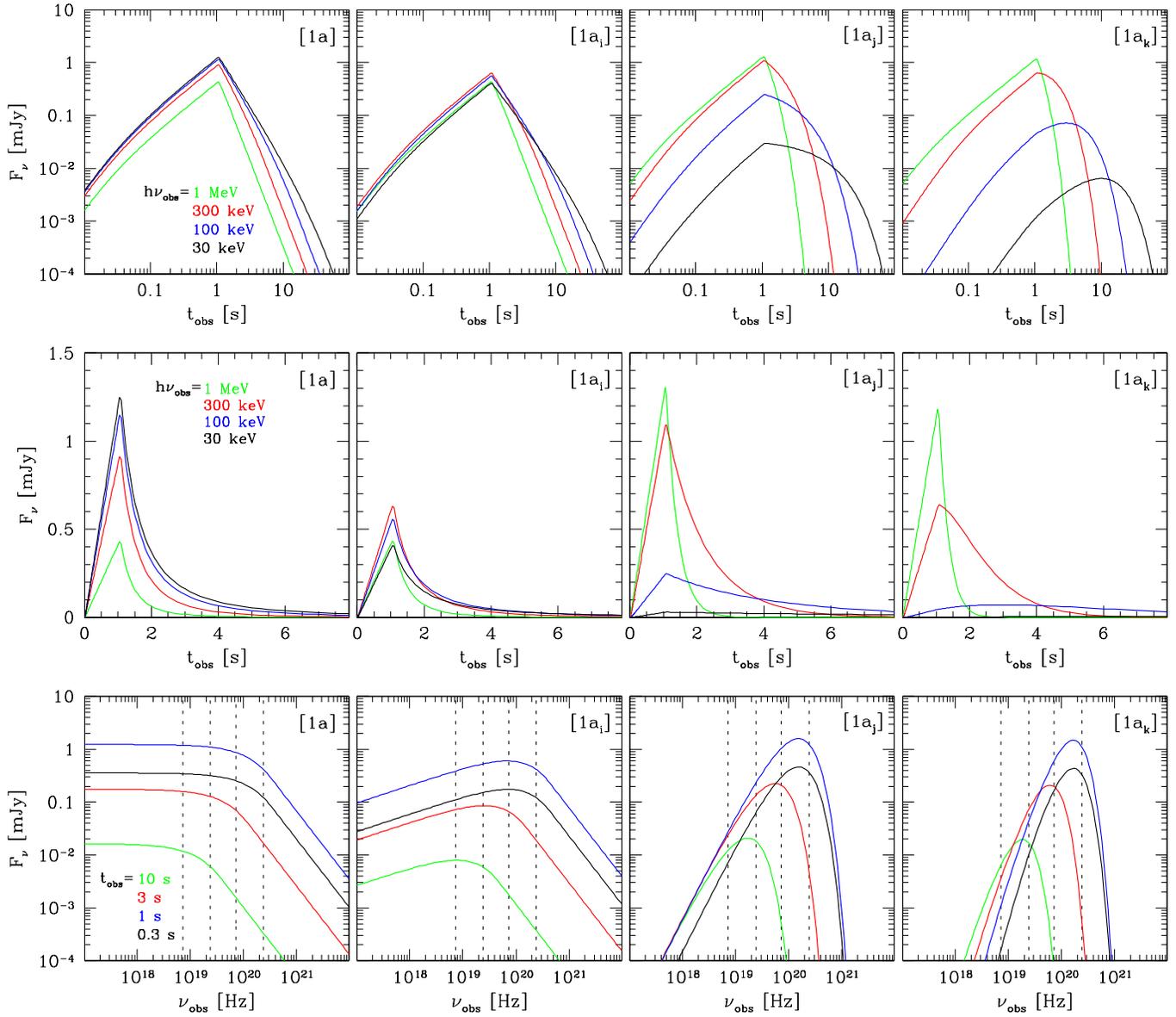}
\caption{
Observed spectral flux $F_{\nu}$ emitted from a relativistic spherical surface that move at a constant Lorentz factor $\Gamma=300$. The emission of the spherical shell is turned off at radius $r_{\rm off} = 3 \times 10^{15}$ cm, which corresponds to the turn-off time at about $\tobs=1.1$ s. Top panels show the light-curves in logarithmic scales at four different energies, 30 keV (black), 100 keV (blue), 300 keV (red), and 1 MeV (green), respectively, while the middle panels show the same four light-curves in linear scales. Bottom panels show the observed spectra at four different observer times, 0.3 s (black), 1 s (blue), 3 s (red), and 10 s (green), respectively. Hence, the blue spectra are dominated by the line-of-sight emission emitted shortly before the turn-off radius, and the red and green spectra are produced purely by the high-latitude emission. The four dotted vertical lines in the bottom panels correspond to the energies, 30 keV, 100 keV, 300 keV, and 1 MeV, respectively, for which the light-curves are calculated. We present four different models here with regard to the shape of the photon spectrum in the co-moving frame, as delineated by the functional form $H(x)$. In the first column (model [1a]), the $H(x)$ is assumed to have a Band-function shape with the typical observed values $\alpha_{\rm B}=-1$ and $\beta_{\rm B}=-2.3$. In the second column (model [1a$_i$]), we keep everything the same as in model [1a], but adopt $\alpha_{\rm B}=-2/3$. In the third column (model [1a$_j$]), we consider the case where the photon spectrum in the co-moving frame assumes a Planck-function shape, i.e., $H(x)=x^3/(e^x-1)$. In the fourth column (model [1a$_k$]), we adopt $H(x)=x^5/(e^x-1)$, although un-physical. The high-latitude curvature effect cannot produce any spectral lag if the photon spectrum in the co-moving frame takes a shape softer than $H(x)=x^2$ below the peak area. Even if some spectral lags are possible for a shape harder than this critical case $x^2$, the resulting spectral lags are essentially invisible, as shown in linear scales.
}
\label{fig:f1}
\end{center}
\end{figure}

Figure~\ref{fig:f1} shows an example calculation of the observed spectral fluxes in different energy bands, which are emitted from a relativistic spherical shell. A constant Lorentz factor $\Gamma=300$ is adopted, and the emission of the shell is turned on at radius $r_{\rm on} = 10^{14}$ cm. The turn-off radius is taken to be $r_{\rm off} = 3 \times 10^{15}$ cm, so that it corresponds to the turn-off time at about $\tobs=1.1$ s. Notice that one can achieve the same turn-off time with different values of $\Gamma$, $r_{\rm on}$, and $r_{\rm off}$ by simultaneously adjusting them, but our conclusion below still remains unchanged. Beyond the turn-off time $\tobs=1.1$ s, the observed spectral flux is produced purely by the high-latitude emission. The number of the electrons in the shell is assumed to increase at a constant injection rate $R_{\rm inj} = 10^{47}~\mbox{s}^{-1}$. The top panels show the light curves\footnote{A real GRB light-curve is composed of many overlapping pulses. The light curves in our calculations are for one single pulse, which is the radiation unit of GRB emission.} in logarithmic scales at four different energies, 30 keV (black), 100 keV (blue), 300 keV (red), and 1 MeV (green), respectively, while the middle panels show the same four light-curves in linear scales. The bottom panels show the observed spectra at four different observer times, 0.3 s (black), 1 s (blue), 3 s (red), and 10 s (green), respectively. Hence, the 0.3 s (black) and 1 s (blue) spectra are dominated by the line-of-sight emission, while the 3 s (red) and 10 s (green) spectra are made purely by the high-latitude emission. The four dotted vertical lines in the bottom panels correspond to the energies, 30 keV, 100 keV, 300 keV, and 1 MeV, respectively, for which the light curves are calculated.

We present four different models in Figure~\ref{fig:f1} with regard to the shape of the photon spectrum in the co-moving frame, as delineated by the functional form $H(x)$. In the first column (model [1a]), the $H(x)$ is assumed to have a Band-function shape with the typical observed values $\alpha_{\rm B}=-1$ and $\beta_{\rm B}=-2.3$. We choose\footnote{In practice, we do so by making use of the synchrotron formulae, given in Equation (\ref{eq:syn}), with the values $B=30$ G and $\gamma_{\rm ch}=10^5$. We stress, however, that the result presented in this Section holds generically, not subject to the synchrotron radiation.} a certain constant value for each of $\nu_{\rm ch}^{\prime}$ and $P_0^{\prime}$ such that the observed spectra are placed at around 1 MeV with the flux density of about 1 mJy,  or $2.4 \times 10^{-6} ~{\rm erg~cm^{-2}~s^{-1}}$, which corresponds to a relatively bright GRB.
In the second column (model [1a$_i$]), we keep everything the same as in model [1a], but adopt $\alpha_{\rm B}=-2/3$ so that the assumed photon spectrum has the same slope as the synchrotron function \citep{rybicki79} below the peak area. As one can see clearly from the figure, the light curves of these two models at different energies peak at the turn-off time simultaneously and do not exhibit a spectral lag. Just as predicted from the curvature effect theory of the spherical shell, the 3 s (red) and 10 s (green) spectra (made by the high-latitude emission) are placed at a progressively lower energy than the 0.3 s (black) and 1 s (blue) spectra (dominated by the line-of-sight emission), but with a progressively weaker flux density. For a constant value of $\Gamma$, the peak energy $E_p$ and the peak flux density $F_{\nu,p}$ of the observed spectra satisfy the following relation during the high-latitude emission, 
\be
F_{\nu,p} \propto E_p^2,
\ee
since $E_p \propto D$ and $F_{\nu,p} \propto D^2$ \citep{uhm15a} where $D$ is the Doppler factor.

In the third column (model [1a$_j$]), we consider the case where the photon spectrum in the co-moving frame assumes a Planck-function shape, i.e., $H(x)=x^3/(e^x-1)$, and the observed spectra are located at a similar characteristic frequency and flux density. The light curves at different energies still do not exhibit a spectral lag. However, we notice that the 1 s (blue), 3 s (red), and 10 s (green) spectra now appear to be coincident with one another asymptotically at low energies. The 3 s (red) and 10 s (green) spectra are expected to coincide with each other at low energies, since $H(x)=x^2$ is satisfied for $x \ll 1$ and the relation $F_{\nu,p} \propto E_p^2$ holds during the high-latitude emission. Given that these two conditions are satisfied, the 1 s (blue) spectrum also appears to be coincident simply because its time 1 s is shortly before the turn-off time at 1.1 s. This coincidence of the spectra indicates a possibility that the 3 s (red) and 10 s (green) spectra can even outshine the 1 s (blue) spectrum at low energies, if the $H(x)$ assumes a shape harder than $H(x)=x^2$ below the peak area. Thus, in the fourth column (model [1a$_k$]), we adopt $H(x)=x^5/(e^x-1)$, although unphysical, and place the spectra at a similar characteristic frequency and flux density. As one can see, the 3 s (red) and 10 s (green) spectra indeed outshine the 1 s (blue) spectrum at low energies, and some spectral lags are produced between the light curves at different energies. However, due to the significant suppression of the flux density during the high-latitude emission, the light curves at low energies become too faint. Consequently, the produced spectral lags are essentially not noticeable, as shown in linear scales.\footnote{Observationally, the prompt emission light-curves are displayed in linear scales. We adopt the same convention in Figures \ref{fig:f1}-\ref{fig:f3}. For Figure \ref{fig:f1}, since the curvature effect predicts a very steep drop in flux density at low energies, we also present the light curves in log scales (top panels) to discern possible spectral lags.}

We conclude that the high-latitude curvature effect cannot produce any spectral lag if the photon spectrum in the co-moving frame takes a shape softer than $H(x)=x^2$ below the peak area. Even if some spectral lags might be possible for a shape harder than this critical case $x^2$, the resulting spectral lags are essentially not noticeable.

%
%

\section{Origin of Spectral Lags}

We will now provide a simple and natural explanation for the observed spectral lags, considering synchrotron radiation that arises from a relativistic spherical surface. The physical picture and the numerical model adopted in our calculations are the same as described in Section~\ref{section:physical_model}. From the synchrotron theory, we express the characteristic frequency $\nu_{\rm ch}^{\prime}$ of the photon spectrum and the spectral power $P_0^{\prime}$ of each electron as follows \citep{rybicki79}
\be
\label{eq:syn}
\nu_{\rm ch}^{\prime}=\frac{3}{16}\, \frac{q_{\rm e} B}{m_{\rm e} c}\, \gamma_{\rm ch}^2,
\quad
P_0^{\prime} = \frac{3 \sqrt{3}}{32}\, \frac{m_{\rm e} c^2\, \sigma_{\rm T} B}{q_{\rm e}},
\ee
where $m_{\rm e}$ and $q_{\rm e}$ are the electron's mass and charge, respectively, and $\sigma_{\rm T}$ is the Thomson cross section. The magnetic field strength $B$ and the characteristic Lorentz factor $\gamma_{\rm ch}$ of the electrons are measured in the co-moving frame of the spherical shell.

\subsection{Magnetic field strength in the emitting region should decrease with radius}

As the spherical shell expands in a 3 dimensional space, the magnetic field strength $B$ in the shell cannot be preserved as a constant; rather, a simple consideration of flux conservation of the magnetic fields yields $B \propto r^{-1}$ for the toroidal component \citep[e.g.][]{spruit01}. Also, possible dissipation of magnetic field energy in the shell (e.g., via reconnection of magnetic field lines) would result in a faster decrease than indicated by the flux conservation. Hence, we consider
\be
\label{eq:B_decrease}
B(r) = B_0 (r/r_0)^{-b},
\ee
where $B_0$ is a normalization value at radius $r_0$, and the index $b \geq 1$. As we recently showed, this decrease of magnetic field strength in the emitting region is an essential element in reproducing the observed shape of prompt emission spectra of GRBs \citep{uhm14}.

Figure~\ref{fig:f2} shows the observed spectral flux emitted from the spherical shell whose magnetic field strength in the co-moving frame follows Equation (\ref{eq:B_decrease}). As in Figure~\ref{fig:f1}, the shell moves at a constant Lorentz factor $\Gamma=300$, and its emission is turned on at radius $r_{\rm on} = 10^{14}$ cm. The turn-off radius is taken to be larger than before, $r_{\rm off} = 3 \times 10^{16}$ cm, so that it corresponds to a larger turn-off time at about $\tobs=11$ s. By choosing so, we can focus on the line-of-sight emission for the timescale of seconds, since the high-latitude curvature effect does not play a significant role when $\tobs<11$ s.
The number of the electrons in the shell is assumed to increase at a constant injection rate $R_{\rm inj} = 10^{47}~\mbox{s}^{-1}$. The top panels show the light curves in linear scales at four different energies, 30 keV (black), 100 keV (blue), 300 keV (red), and 1 MeV (green), respectively, while the bottom panels show the observed spectra at four different observer times, 0.3 s (black), 1 s (blue), 3 s (red), and 10 s (green), respectively. Therefore, all four spectra shown here are dominated by the line-of-sight emission. The four dotted vertical lines in the bottom panels correspond to the energies, 30 keV, 100 keV, 300 keV, and 1 MeV, respectively, for which the light curves are calculated.

In Figure~\ref{fig:f2}, we adopt a Band-function shape with the parameters $\alpha_{\rm B}=-0.8$ and $\beta_{\rm B}=-2.3$, as for the shape of the photon spectrum in the co-moving frame. Based on the typical parameters of GRB prompt emission \citep[e.g.][]{zhangyan11,uhm14}, we choose the values $B_0=30$ G at $r_0=10^{15}$ cm, and $\gamma_{\rm ch}=8 \times 10^4$, which locate the observed spectra at around the observing energy bands. In the first column (model [1b]) we take $b=1$ (i.e., flux-conservation case), while in the second column (model [1c]) we take $b=1.25$. The third column (model [1d]) takes $b=1.5$.

Due to a decreasing strength of magnetic field in the emitting region, the characteristic frequency $\nu_{\rm ch}^{\prime}$ of the photon spectrum in the co-moving frame, given in Equation (\ref{eq:syn}), decreases in time. Consequently, the observed spectra sweep through the observing energy bands as the observer time $\tobs$ elapses. The larger the index $b$ is, the faster the sweep-through is. The observed flux-level can also decrease in time, due to a decreasing spectral power $P_0^{\prime}$ of each electron in the co-moving frame, even though the number of radiating electrons $N$ in the shell increases in time. As one can see from the figure, some spectral lags are produced in our model light-curves. We stress that these spectral lags occur well before the turn-off time, i.e., well before the high-latitude emission dominates over the line-of-sight emission.

Although encouraging, the models here with a decreasing strength of magnetic field in the emitting region alone cannot explain the observations. In Section~\ref{section:observed_properties}, we will show that the details of the model light-curves shown in Figure~\ref{fig:f2} are not consistent with the observed properties of spectral lags. In particular, the pulse profiles of these models are too broad at low frequencies, leading to a too steep slope in the $t_p- \nu_{\rm obs}$ and width$- \nu_{\rm obs}$ relations.

We conclude this Section by making an important point that {\em in order to produce some spectral lags, the observed spectrum has to be curved, and the spectral peak has to move from high-energy to low-energy as a function of time}. Imagine that the peak energy of a curved photon spectrum remains constant during a time span of observation. Then, moving the spectrum upward or downward would only result in a simultaneous rise or fall in the light curves at different energies, without making any spectral lag.

\subsection{Emitting region itself should undergo rapid bulk acceleration}

From the observations of the steep-decay phase of GRB X-ray flares \citep{burrows05}, we recently found clear observational evidence that the X-ray flare emitting region undergoes rapid bulk acceleration as the photons are emitted, hence providing a signature for a Poynting-flux-dominated outflow in the emitting region of X-ray flares \citep{uhm15b}. The finding applies to the majority of X-ray flares detected by Swift, suggesting that bulk acceleration may be ubiquitous \citep{jia15}. Recalling that the X-ray flares share a similar physical mechanism as the prompt emission itself, we investigate here the effect of bulk acceleration on the formation of spectral lags in prompt emission. We consider
\be
\label{eq:bulk_gamma}
\Gamma(r) = \Gamma_0 (r/r_0)^{s},
\ee
where $\Gamma_0$ is a normalization value at radius $r_0$, and the index $s$ describes the degree of bulk acceleration.

\begin{figure}
\begin{center}
\includegraphics[width=18cm]{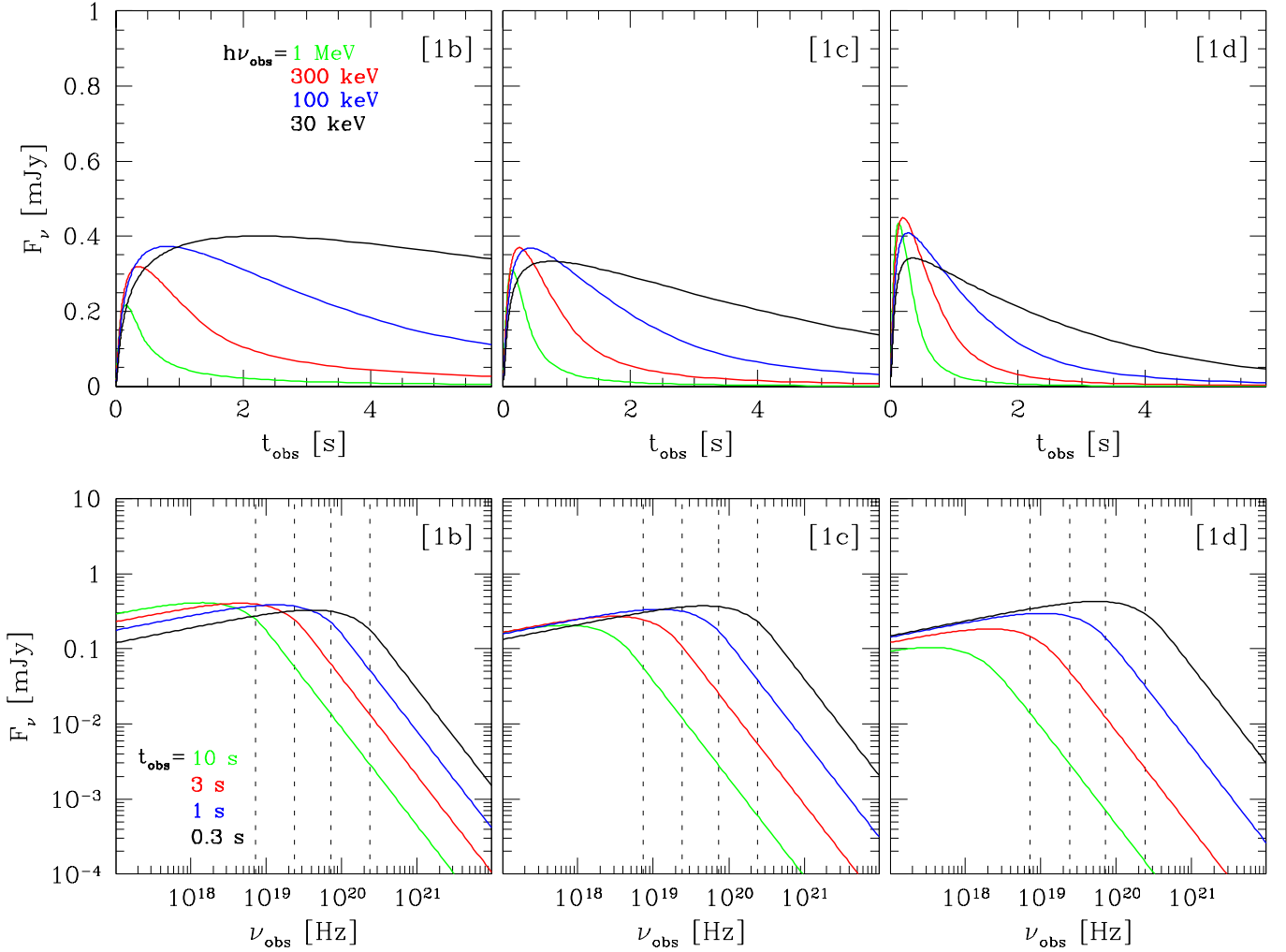}
\caption{
Observed spectral flux $F_{\nu}$ emitted from a relativistic spherical shell whose magnetic field strength $B$ in the co-moving frame decreases in radius $r$, following $B(r) \propto r^{-b}$. The shell still moves at a constant Lorentz factor $\Gamma=300$. Top panels show the light curves in linear scales at four different energies, 30 keV (black), 100 keV (blue), 300 keV (red), and 1 MeV (green), respectively, while the bottom panels show the observed spectra at four different observer times, 0.3 s (black), 1 s (blue), 3 s (red), and 10 s (green), respectively. All four spectra shown here are dominated by the line-of-sight emission. The four dotted vertical lines in the bottom panels correspond to the energies, 30 keV, 100 keV, 300 keV, and 1 MeV, respectively, for which the light curves are calculated. As for the shape of the photon spectrum in the co-moving frame, we adopt a Band-function shape with the parameters $\alpha_{\rm B}=-0.8$ and $\beta_{\rm B}=-2.3$. In the first column (model [1b]) we take $b=1$ (i.e., flux-conservation case), while in the second column (model [1c]) we take $b=1.25$. The third column (model [1d]) takes $b=1.5$. Due to a decreasing strength of magnetic field in the emitting region, the observed spectra sweep through the observing energy bands as the observer time $\tobs$ elapses. Some spectral lags are produced in the model light-curves, but are not consistent with the observed properties of spectral lags; see Section~\ref{section:observed_properties}.
}
\label{fig:f2}
\end{center}
\end{figure}

\begin{figure}
\begin{center}
\includegraphics[width=18cm]{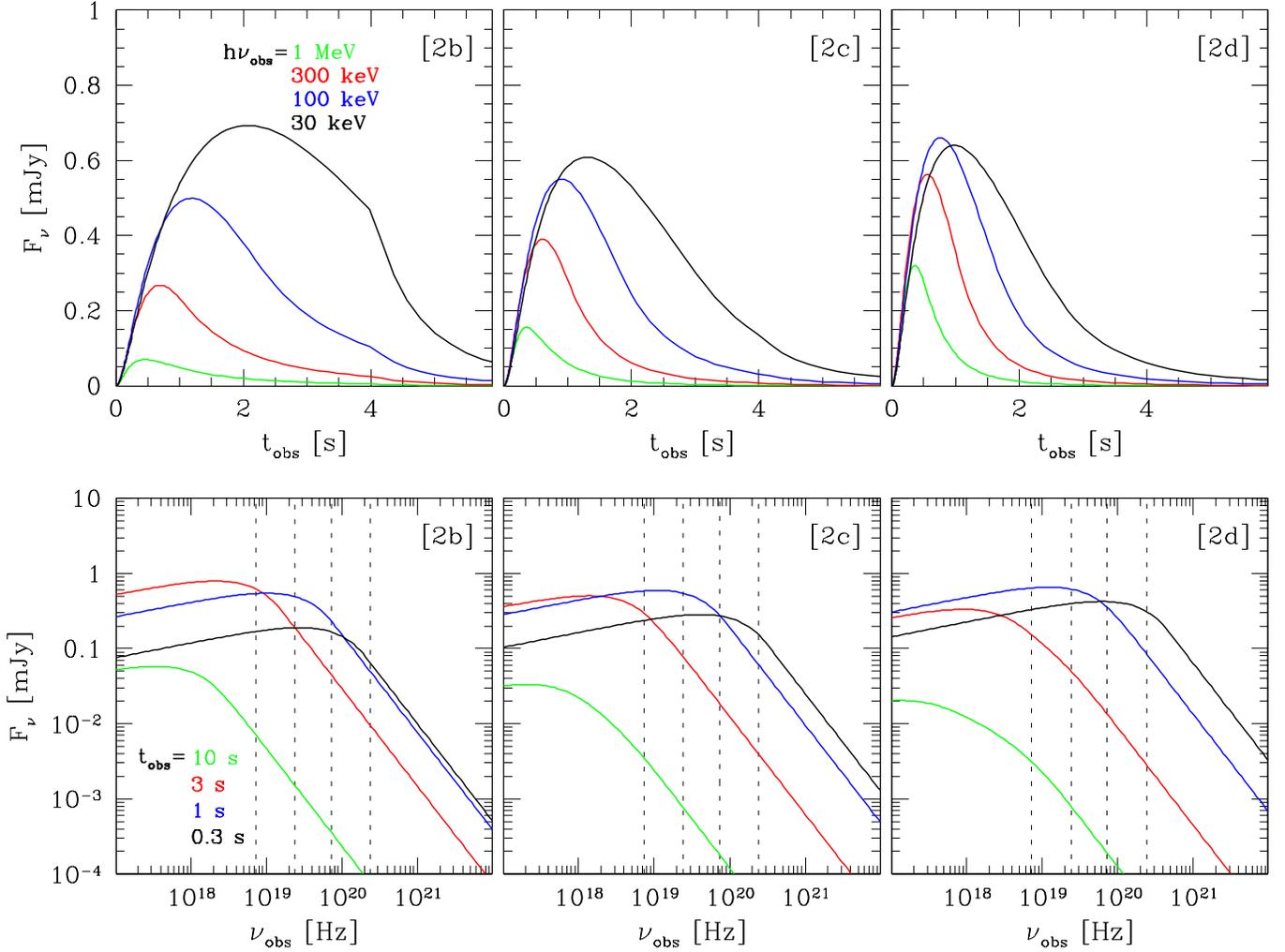}
\caption{
Observed spectral flux $F_{\nu}$ emitted from a relativistic spherical shell whose bulk Lorentz factor follows $\Gamma(r) \propto r^s$ and whose magnetic field strength in the co-moving frame follows $B(r) \propto r^{-b}$. The index $s$ describes the degree of bulk acceleration, and is taken to be $s = 0.35$. Shown in the first column is the model [2b] with $b=1$. The second column (model [2c]) takes $b=1.25$, while the third column (model [2d]) takes $b=1.5$. Top panels show the light curves at four different energies, 30 keV (black), 100 keV (blue), 300 keV (red), and 1 MeV (green), respectively, while the bottom panels show the observed spectra at four different observer times, 0.3 s (black), 1 s (blue), 3 s (red), and 10 s (green), respectively. The light curves exhibit a clear pattern of spectral lags. The details of the light curves in models [2c] and [2d] are in a good agreement with the observed properties of spectral lags; see Section~\ref{section:observed_properties}. The combined effect of an accelerating profile of $\Gamma(r)$ and a decreasing profile of $B(r)$ allows for a successful reproduction of the observed spectral lags, without a need to invoke for the high-latitude curvature effect.
}
\label{fig:f3}
\end{center}
\end{figure}

We begin with the models [1b], [1c], and [1d] (presented in Figure~\ref{fig:f2} for a decreasing profile of $B(r)$ in the emitting region) and let these models undergo bulk acceleration instead of coasting with $\Gamma(r)=300$. Since no first-principle prediction is available to quantify the acceleration of the jet, we explore different acceleration profiles $\Gamma(r)$ to reach a satisfactory match with the data. The characteristic electron Lorentz factor $\gamma_{\rm ch}$ is correspondingly adjusted to assure that the observed $E_p$ is around 1 MeV. One example calculation adopts $\Gamma_0 = 250$ at $r_0 = 10^{15}$ cm and $s = 0.35$, and $\gamma_{\rm ch}=5 \times 10^4$, with all other model parameters the same as in Figure~\ref{fig:f2}. The results are shown in Figure~\ref{fig:f3}.

Shown in the first column in Figure~\ref{fig:f3} is the model [2b] with $b=1$ as for the model [1b]. The second column (model [2c]) takes $b=1.25$ as in the model [1c], while the third column (model [2d]) is with $b=1.5$ as in the model [1d]. Again, the top panels show the light curves at four different energies, 30 keV (black), 100 keV (blue), 300 keV (red), and 1 MeV (green), respectively, while the bottom panels show the observed spectra at four different observer times, 0.3 s (black), 1 s (blue), 3 s (red), and 10 s (green), respectively. We kept the same turn-on and turn-off radius here as in Figure~\ref{fig:f2}, but for this accelerating profile of $\Gamma(r)$, the turn-off point corresponds to the turn-off time at about $\tobs = 4.0$ s in Figure~\ref{fig:f3}. Hence, the 10 s (green) spectra are made purely by the high-latitude emission.

As one can see from the figure, the light curves now exhibit a clear pattern of spectral lags, successfully reproducing the observations. In Section~\ref{section:observed_properties}, we will show that the details of the light curves in models [2c] and [2d] are indeed in a good agreement with the observed properties of spectral lags.

Due to a decreasing profile of $B(r)$ in the emitting region, the observed spectra sweep through the observing energy bands again as the observer time $\tobs$ elapses. The larger the index $b$, the faster the sweep-through. However, the sweep-through in Figure~\ref{fig:f3} occurs in a more complicated way than in Figure~\ref{fig:f2}, since the effect of bulk acceleration is added. When an accelerating profile of $\Gamma(r)$ is combined with a decreasing profile of $B(r)$, the observed flux-level can exhibit an initial rise followed by a fall at a later time. This behavior reshapes our model light-curves in such a clever way that their properties come close to the observations.

This complicated behavior may be understood by the following rough analytical estimates. For the observed spectra dominated by the line-of-sight emission, we notice that the peak energy $E_p \propto \Gamma \nu'_{\rm ch} \propto \Gamma B$, whereas the peak flux density $F_{\nu,p} \propto N \Gamma B$ with the total number of electrons $N \propto t' \propto r/\Gamma$ for constant injection rate in the co-moving frame. Hence, one roughly has
\begin{equation}
 E_p \propto r^{s-b} \propto t_{\rm obs}^{\frac{s-b}{1-2s}}, ~~~ F_{\nu,p} \propto r^{1-b} \propto t_{\rm obs}^{\frac{1-b}{1-2s}},
\end{equation}
for $s < 1/2$. One can see that these two critical parameters to calculate the light curves at different energy bands both depend on the $b$ and $s$ parameters, but in different ways. Introducing bulk acceleration ($s$ parameter) is essential to reproduce the correct pulse broadness and its energy evolution as observed.

We also remark that the turning-off of the shell's emission, visible at about $\tobs = 4.0$ s in the model [2b], is nearly invisible in the models [2c] and [2d] so that the high-latitude emission can take over without being noticed. This is also an attractive aspect of the models presented here. We conclude that the combined effect of an accelerating profile of $\Gamma(r)$ and a decreasing profile of $B(r)$ allows for a successful reproduction of the observed spectral lags, well before the shell's emission is turned off, i.e., without a need to invoke for the high-latitude curvature effect.

\subsection{Comparing to the observed properties of spectral lags} \label{section:observed_properties}

\begin{figure}
\begin{center}
\includegraphics[width=11cm]{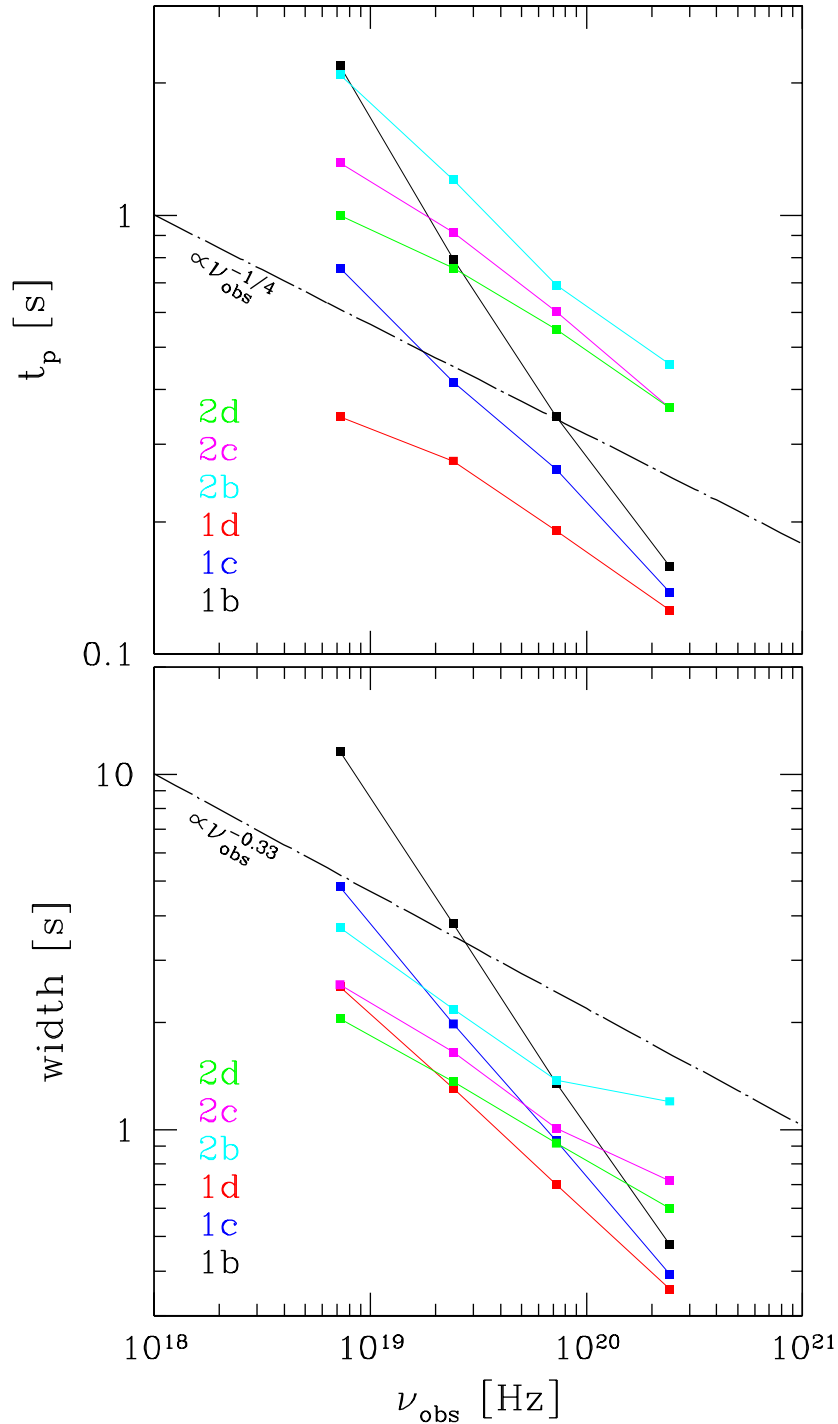}
\caption{
Properties of the model light-curves as compared to the observations of spectral lags. For the six models [1b], [1c], [1d], [2b], [2c], and [2d], we plot the peak time $t_p$ (top panel) and the width (bottom panel) of each light curve against the observing frequency $\nuobs$ of that light curve. The dot-dashed lines indicate the relations $t_p \propto \nuobs^{-1/4}$ (top panel) and $width \propto \nuobs^{-0.33}$ (bottom panel), respectively, which represent the observed properties of spectral lags \citep{norris96,liang06}. Only slopes of the relations in log-log scales are meaningful, so the normalizations of the two dot-dashed lines are chosen arbitrarily. It is clear that the models [1b], [1c], and [1d] are not consistent with the observations, since the slopes of at least one relation are too steep. The model [2b] is better, although the first relation is somewhat too steep. On the other hand, the models [2c] and [2d] are in a good agreement with the observed properties of spectral lags.
}
\label{fig:f4}
\end{center}
\end{figure}

Now we compare the properties of our model light-curves to the observations. We first take the model [1b] and find the peak time $t_p$ and the width individually for each of the model's four light curves. The peak time is the observer time where the light curve peaks, and the width is defined as the full observer-time interval that covers the half of the maximum flux-density of the light curve. In Figure~\ref{fig:f4}, we plot the peak time (top panel) and the width (bottom panel) of each light curve against the observing frequency of that light curve, and then connect the plotted points with a line. We repeat this for all of the six models [1b], [1c], [1d], [2b], [2c], and [2d], and show the results together in Figure~\ref{fig:f4}, for comparison.

The two dot-dashed lines in Figure~\ref{fig:f4} indicate the relations $t_p \propto \nuobs^{-1/4}$ (top panel) and $width \propto \nuobs^{-0.33}$ (bottom panel), respectively, which represent the observed properties of spectral lags \citep{norris96,liang06}. Notice that only the slopes of the two relations in log-log scales give meaningful constraints on the models. Thus, in Figure~\ref{fig:f4}, the two relations are shown with arbitrary normalization and compared with our model results. It is clear that the models [1b], [1c], and [1d] are not consistent with the observations, since they have too steep slopes in at least one of the two relations. The model [2b] is better, but the $t_p-\nu_{\rm obs}$ slope is still somewhat steeper than the observed relation. On the other hand, the models [2c] and [2d] are in a good agreement with the observed properties of spectral lags.

%
%

\section{Conclusions and Discussion}

In this paper, based on a simple but general emission model and some observational constraints (e.g., the duration of GRB pulses and the statistical properties of GRB spectral lags), we explore model parameters that are suitable to reproduce the observations. Since the spectral lags sample the broad pulses rather than the rapid variabilities, we use the duration of broad pulses to constrain the radius of the emission region from the central engine. We showed that the high-latitude curvature effect, which has been widely considered in previous studies, cannot reproduce the observed spectral lags. In particular, if the emission spectrum is broad (such as the Band function as observed), the curvature effect cannot give rise to any spectral lags. Some spectral lags may be produced if the spectrum is un-physically narrow (harder than $x^2$ below the peak). However, the resulting spectral lags are essentially not noticeable, since the flux-density levels at different frequencies have large contrasts, inconsistent with the data.
Within the framework of an optically-thin synchrotron radiation model (supported by the required large radius), we further derive the physical conditions required to reproduce the observed spectral lags. Based on the observed $E_p$ and $\Gamma$ for typical GRBs, we derive the relevant model parameters such as $B_0$, $\gamma_{\rm ch}$, etc, and calculate the single-pulse light curves at different energy bands.

We find that in order to reproduce the observed spectral lags, a physical model needs to satisfy the following constraints:
\begin{itemize}
 \item The emission radius has to be large (above several $10^{14}$ cm from the central engine), in the optically thin region well above the photosphere. This is required by the observed duration of the slow-component pulses and the typical Lorentz factor inferred from GRB observations (Equation (\ref{eq:r}));
 \item The $\gamma$-ray photon spectrum should be curved (as observed);
 \item The magnetic field strength in the emitting region should decrease with radius, which is consistent with an expanding jet;
 \item The emission region itself should undergo rapid bulk acceleration as the prompt emission is produced.
\end{itemize}

The last requirement (bulk acceleration) poses interesting constraints on the jet composition of GRBs. Within the traditional fireball model, bulk acceleration is thermally driven, and proceeded at a much smaller radius \citep{meszaros93,piran93,kobayashi99}. At the large emission radius required by the observations, the thermal energy is essentially converted to kinetic energy of the outflow, and the significant bulk acceleration is impossible. If the outflow is Poynting-flux-dominated \citep{granot11}, or hybrid with a significant Poynting flux component \citep{gaozhang15}, the outflow keeps accelerating in a much longer distance scale, as the Poynting flux energy is gradually converted to kinetic energy. If in the emission region the outflow is still moderately magnetized, as envisaged in the ICMART model of GRBs \citep{zhangyan11,deng15}, additional bulk acceleration is expected as collision-induced magnetic reconnection efficiently converts magnetic energy to particle energy and radiation. The required bulk acceleration in the prompt emission region is therefore consistent with such a scenario, suggesting that the ICMART-like mechanism is the mechanism to produce GRB prompt emission.

We note that based on a completely different argument (the decay slope of X-ray flare tails), we also drew the conclusion that the emission regions of essentially all the X-ray flares also undergo significant bulk acceleration \citep{uhm15b,jia15}, suggesting that a moderate Poynting-flux is likely ubiquitous among GRBs and their subsequent X-ray flares.

%
%

\acknowledgments
We thank Chuck Dermer and the anonymous referee for helpful comments. This work is supported by NASA through an Astrophysical Theory Program (grant number NNX 15AK85G) and an Astrophysics Data Analysis Program (grant number NNX 14AF85G).

%
%


\begin{thebibliography}{44}
\expandafter\ifx\csname natexlab\endcsname\relax\def\natexlab#1{#1}\fi

\bibitem[{{Ahlgren} {et~al.}(2015){Ahlgren}, {Larsson}, {Nymark}, {Ryde}, \&
  {Pe'er}}]{ahlgren15}
{Ahlgren}, B., {Larsson}, J., {Nymark}, T., {Ryde}, F., \& {Pe'er}, A. 2015,
  \mnras, 454, L31

\bibitem[{{Band} {et~al.}(1993){Band}, {Matteson}, {Ford}, {Schaefer},
  {Palmer}, {Teegarden}, {Cline}, {Briggs}, {Paciesas}, {Pendleton}, {Fishman},
  {Kouveliotou}, {Meegan}, {Wilson}, \& {Lestrade}}]{band93}
{Band}, D., {Matteson}, J., {Ford}, L., {et~al.} 1993, \apj, 413, 281

\bibitem[{{Beloborodov}(2010)}]{beloborodov10}
{Beloborodov}, A.~M. 2010, \mnras, 407, 1033

\bibitem[{{Burgess} {et~al.}(2014){Burgess}, {Preece}, {Ryde}, {Veres},
  {M{\'e}sz{\'a}ros}, {Connaughton}, {Briggs}, {Pe'er}, {Iyyani}, {Goldstein},
  {Axelsson}, {Baring}, {Bhat}, {Byrne}, {Fitzpatrick}, {Foley}, {Kocevski},
  {Omodei}, {Paciesas}, {Pelassa}, {Kouveliotou}, {Xiong}, {Yu}, {Zhang}, \&
  {Zhu}}]{burgess14}
{Burgess}, J.~M., {Preece}, R.~D., {Ryde}, F., {et~al.} 2014, \apjl, 784, L43

\bibitem[{{Burrows} {et~al.}(2005){Burrows}, {Romano}, {Falcone}, {Kobayashi},
  {Zhang}, {Moretti}, {O'Brien}, {Goad}, {Campana}, {Page}, {Angelini},
  {Barthelmy}, {Beardmore}, {Capalbi}, {Chincarini}, {Cummings}, {Cusumano},
  {Fox}, {Giommi}, {Hill}, {Kennea}, {Krimm}, {Mangano}, {Marshall},
  {M{\'e}sz{\'a}ros}, {Morris}, {Nousek}, {Osborne}, {Pagani}, {Perri},
  {Tagliaferri}, {Wells}, {Woosley}, \& {Gehrels}}]{burrows05}
{Burrows}, D.~N., {Romano}, P., {Falcone}, A., {et~al.} 2005, Science, 309,
  1833

\bibitem[{{Daigne} {et~al.}(2011){Daigne}, {Bo{\v s}njak}, \&
  {Dubus}}]{daigne11}
{Daigne}, F., {Bo{\v s}njak}, {\v Z}., \& {Dubus}, G. 2011, \aap, 526, A110

\bibitem[{{Deng} {et~al.}(2015){Deng}, {Li}, {Zhang}, \& {Li}}]{deng15}
{Deng}, W., {Li}, H., {Zhang}, B., \& {Li}, S. 2015, \apj, 805, 163

\bibitem[{{Dermer}(2004)}]{dermer04}
{Dermer}, C.~D. 2004, \apj, 614, 284

\bibitem[{{Drenkhahn} \& {Spruit}(2002)}]{drenkhahn02}
{Drenkhahn}, G., \& {Spruit}, H.~C. 2002, \aap, 391, 1141

\bibitem[{{Fishman} \& {Meegan}(1995)}]{fishman95}
{Fishman}, G.~J., \& {Meegan}, C.~A. 1995, \araa, 33, 415

\bibitem[{{Gao} \& {Zhang}(2015)}]{gaozhang15}
{Gao}, H., \& {Zhang}, B. 2015, \apj, 801, 103

\bibitem[{{Gao} {et~al.}(2012){Gao}, {Zhang}, \& {Zhang}}]{gao12}
{Gao}, H., {Zhang}, B.-B., \& {Zhang}, B. 2012, \apj, 748, 134

\bibitem[{{Giannios}(2008)}]{giannios08}
{Giannios}, D. 2008, \aap, 480, 305

\bibitem[{{Goodman}(1986)}]{goodman86}
{Goodman}, J. 1986, \apjl, 308, L47

\bibitem[{{Granot} {et~al.}(2011){Granot}, {Komissarov}, \&
  {Spitkovsky}}]{granot11}
{Granot}, J., {Komissarov}, S.~S., \& {Spitkovsky}, A. 2011, \mnras, 411, 1323

\bibitem[{{Hakkila} \& {Preece}(2011)}]{hakkila11}
{Hakkila}, J., \& {Preece}, R.~D. 2011, \apj, 740, 104

\bibitem[{{Hasco{\"e}t} {et~al.}(2012){Hasco{\"e}t}, {Daigne}, \&
  {Mochkovitch}}]{hascoet12}
{Hasco{\"e}t}, R., {Daigne}, F., \& {Mochkovitch}, R. 2012, \aap, 542, L29

\bibitem[{{Ioka} \& {Nakamura}(2001)}]{ioka01}
{Ioka}, K., \& {Nakamura}, T. 2001, \apjl, 554, L163

\bibitem[{{Jia} {et~al.}(2015)}]{jia15}
{Jia}, L.-W., {Uhm}, Z. L. \& {Zhang}, B. 2015, \apjs, submitted (arXiv:1509.04871)

\bibitem[{{Kobayashi} {et~al.}(1999){Kobayashi}, {Piran}, \&
  {Sari}}]{kobayashi99}
{Kobayashi}, S., {Piran}, T., \& {Sari}, R. 1999, \apj, 513, 669

\bibitem[{{Kumar} \& {Zhang}(2015)}]{kumarzhang15}
{Kumar}, P., \& {Zhang}, B. 2015, \physrep, 561, 1

\bibitem[{{Lazzati} \& {Begelman}(2010)}]{lazzati10}
{Lazzati}, D., \& {Begelman}, M.~C. 2010, \apj, 725, 1137

\bibitem[{{Liang} {et~al.}(2006){Liang}, {Zhang}, {O'Brien}, {Willingale},
  {Angelini}, {Burrows}, {Campana}, {Chincarini}, {Falcone}, {Gehrels}, {Goad},
  {Grupe}, {Kobayashi}, {M{\'e}sz{\'a}ros}, {Nousek}, {Osborne}, {Page}, \&
  {Tagliaferri}}]{liang06}
{Liang}, E.~W., {Zhang}, B., {O'Brien}, P.~T., {et~al.} 2006, \apj, 646, 351

\bibitem[{{Lu} {et~al.}(2012){Lu}, {Wei}, {Liang}, {Zhang}, {L{\"u}}, {L{\"u}},
  {Lei}, \& {Zhang}}]{lu12}
{Lu}, R.-J., {Wei}, J.-J., {Liang}, E.-W., {et~al.} 2012, \apj, 756, 112

\bibitem[{Lu} {et~al.}(2016)]{lu15} Lu, R. J., et al. 2016, in preparation

\bibitem[{{Lundman} {et~al.}(2013){Lundman}, {Pe'er}, \& {Ryde}}]{lundman13}
{Lundman}, C., {Pe'er}, A., \& {Ryde}, F. 2013, \mnras, 428, 2430

\bibitem[{{M\'esz\'aros} {et~al.}(1993){M\'esz\'aros}, {Laguna}, \&
  {Rees}}]{meszaros93}
{M\'esz\'aros}, P., {Laguna}, P., \& {Rees}, M.~J. 1993, \apj, 415, 181

\bibitem[{{M{\'e}sz{\'a}ros} \& {Rees}(2000)}]{meszarosrees00}
{M{\'e}sz{\'a}ros}, P., \& {Rees}, M.~J. 2000, \apj, 530, 292

\bibitem[{{Norris}(2002)}]{norris02}
{Norris}, J.~P. 2002, \apj, 579, 386

\bibitem[{{Norris} {et~al.}(2000){Norris}, {Marani}, \& {Bonnell}}]{norris00}
{Norris}, J.~P., {Marani}, G.~F., \& {Bonnell}, J.~T. 2000, \apj, 534, 248

\bibitem[{{Norris} {et~al.}(1996){Norris}, {Nemiroff}, {Bonnell}, {Scargle},
  {Kouveliotou}, {Paciesas}, {Meegan}, \& {Fishman}}]{norris96}
{Norris}, J.~P., {Nemiroff}, R.~J., {Bonnell}, J.~T., {et~al.} 1996, \apj, 459,
  393

\bibitem[{{Pacz\'ynski}(1986)}]{paczynski86}
{Pacz\'ynski}, B. 1986, \apjl, 308, L43

\bibitem[{{Pe'er}(2015)}]{peer15}
{Pe'er}, A. 2015, Advances in Astronomy, 2015, 22

\bibitem[{{Pe'er} {et~al.}(2006){Pe'er}, {M{\'e}sz{\'a}ros}, \&
  {Rees}}]{peer06}
{Pe'er}, A., {M{\'e}sz{\'a}ros}, P., \& {Rees}, M.~J. 2006, \apj, 642, 995

\bibitem[{{Piran} {et~al.}(1993){Piran}, {Shemi}, \& {Narayan}}]{piran93}
{Piran}, T., {Shemi}, A., \& {Narayan}, R. 1993, \mnras, 263, 861

\bibitem[{{Rees} \& {M\'esz\'aros}(1994)}]{rees94}
{Rees}, M.~J., \& {M\'esz\'aros}, P. 1994, \apjl, 430, L93

\bibitem[{{Rees} \& {M{\'e}sz{\'a}ros}(2005)}]{rees05}
{Rees}, M.~J., \& {M{\'e}sz{\'a}ros}, P. 2005, \apj, 628, 847

\bibitem[{{Rybicki} \& {Lightman}(1979)}]{rybicki79}
{Rybicki}, G.~B., \& {Lightman}, A.~P. 1979, {Radiative processes in
  astrophysics} (New York, Wiley-Interscience, 1979.~393 p.)

\bibitem[{{Shemi} \& {Piran}(1990)}]{shemi90}
{Shemi}, A., \& {Piran}, T. 1990, \apjl, 365, L55

\bibitem[{{Shen} {et~al.}(2005){Shen}, {Song}, \& {Li}}]{shen05}
{Shen}, R.-F., {Song}, L.-M., \& {Li}, Z. 2005, \mnras, 362, 59

\bibitem[{{Shenoy} {et~al.}(2013)}]{shenoy13}
{Shenoy}, A., {Sonbas}, E., {Dermer}, C. et al., 2013, \apj, 778, 3

\bibitem[{{Spruit} {et~al.}(2001){Spruit}, {Daigne}, \& {Drenkhahn}}]{spruit01}
{Spruit}, H. C., {Daigne}, F., \& {Drenkhahn}, G. 2001, \aap, 369, 694


\bibitem[{{Thompson}(2006)}]{thompson06}
{Thompson}, C. 2006, \apj, 651, 333

\bibitem[{{Uhm} \& {Zhang}(2014)}]{uhm14}
{Uhm}, Z.~L., \& {Zhang}, B. 2014, Nature Physics, 10, 351

\bibitem[{{Uhm} \& {Zhang}(2015a)}]{uhm15a}
---. 2015a, \apj, 808, 33

\bibitem[{{Uhm} \& {Zhang}(2015b)}]{uhm15b}
---. 2015b, arXiv:1509.03296

\bibitem[{{Ukwatta} {et~al.}(2012){Ukwatta}}]{ukwatta12}
{Ukwatta}, T. N., et al. 2012, \mnras, 419, 614

\bibitem[{{Vurm} {et~al.}(2011){Vurm}, {Beloborodov}, \& {Poutanen}}]{vurm11}
{Vurm}, I., {Beloborodov}, A.~M., \& {Poutanen}, J. 2011, \apj, 738, 77

\bibitem[{{Zhang} \& {Yan}(2011)}]{zhangyan11}
{Zhang}, B., \& {Yan}, H. 2011, \apj, 726, 90

\bibitem[{{Zhang} \& {Zhang}(2014)}]{zhangzhang14}
{Zhang}, B., \& {Zhang}, B. 2014, \apj, 782, 92

\bibitem[{Zhang} {et~al.}(2009)]{zhang09}
{Zhang}, B., {Zhang}, B.-B., Virgili, F. J., Liang, E. W., Kann, D. A. et al. 2009, \apj, 703, 1696

\bibitem[{Zhang} {et~al.}(2016)]{zhangbb15}
{Zhang}, B.-B., Uhm, Z. L., Connaughton, V., Briggs, M. S., \& Zhang, B. 2016, \apj, 816, 72

\end{thebibliography}

%
%

\newpage

\end{document}